\documentstyle[aps,prl,multicol]{revtex}
\begin{document}
%%%%%%%%%%%%%%%%%%%%%%%%%%%%%%%%%%%%%%
\def\ibd {{\it ibid. }}
\def\bfj{{\bf j}}
\def\bfq{{\bf q}}
\def\bfk{{\bf k}}
\def\bfrho{{\bf \rho}}
\def\om{\omega}
%%%%%%%%%%%%%%%%%%%%%%%%%%%%%%%%%%%%%%%
\preprint{CNAT-99-2}
\title{Two-Dimensional Gross-Pitaevskii Equation: 
Theory of Bose-Einstein Condensation and the Vortex State}
\author{Sang-Hoon Kim,$^1$  Changyeon Won,$^2$ Sung Dahm Oh,$^3$
 and Wonho Jhe $^{2,\dag}$}
\address{$^1$Division of Liberal Arts, Mokpo National Maritime University,
 Mokpo 530-729, Korea}
\address{$^2$Center for Nearfield Atom-photon Technology
and Physics Department, Seoul National University,
 Seoul 151-742, Korea}
\address{$^3$Physics Department, Sookmyung Women's University, Seoul
 140-742, Korea}
\date{\today}
\maketitle
\begin{abstract}
We derive the Gross-Pitaevskii equation in two-dimension 
from the first principles of two-dimensional scattering theory.
Numerical calculation of the condensate wave function shows that 
atoms in a two-dimensional harmonic trap can be condensed  
into its ground state.   Moreover, the ground state energy 
and the wave function of two-dimensional  vortex state 
are also obtained.   Quantitative comparisons between 2D results 
and 3D ones are made in detail.
\end{abstract}
\draft
\vspace{.25in}
%\newpage
\begin{multicols}{2}
%%%%%%%%%%%%%%%%%%%%%%%%%%%%%%%%%%%%%%%%%%%%%
%%%%%%%   MAIN
%%%%%%%%%%%%%%%%%%%%%%%%%%%%%%%%%%%%%%%%%%%%%
	The recent experimental realization of Bose-Einstein Condensation (BEC) of 
dilute alkali Bose atoms in magnetic traps \cite{ande,brad,davi} has generated
 immense interest and activities in theoretical and experimental physics.
An interesting, but much less addressed, question is whether such a phase 
transition due to global coherence exists in two-dimensional (2D) space.
	 It has been known that  2D Bose atoms do not have a long-range order, 
and consequently BEC of a uniform (untrapped) Bose gas can not 
occur in 2D at a finite temperature since thermal fluctuations 
destabilize the condensate \cite{hohe,kim}.

 		There has been no direct experimental observation of 2D BEC yet,
except the recent related works such as quasi-condensate of atomic
 hydrogen \cite{quasi} and quasi-2D gas of laser cooled atoms \cite{2d}.
Nevertheless,  recent theoretical studies have suggested a possibility of
 BEC in 2D in a trapped condition. However, those
theoretical approaches so far have been only second-handed.
	Tempere and Devreese studied harmonically interacting
bosons in 2D\cite{temp} and calculated the critical temperature by 
thermodynamic arguments to show the possible occurrence of 2D BEC.
However, it could not provide any information on the 2D-condensate wave
 function necessary for the atomic density profile.

		An analytical approach to interacting bosons in a 2D trap by 
Gross-Pitaevskii equation (GPE), a nonlinear Schr\''{o}dinger equation
for the macroscopic wave function of weakly interacting bosons,
 was also attempted by Gonzalez {\it et. al.} \cite{gonz}, 
but they inevitably faced with a dimensional inconsistency 
in the relation between coupling constant and scattering length.
In other words, the well-known 3D result, $U_0 =  4\pi \hbar^2 a /M$ 
(interaction strength $U_0$,  s-wave scattering length  $a$, 
and  atomic mass $M$) is  not directly applicable to 2D GPE.
Bayindir and Tanatar recently studied the 2D GPE within the two-fluid 
model of the mean-field many-body quantum 
statistical theory and could calculate the condensate fraction \cite{bayi}.
However, their results were obtained not from the direct solution 
of the 2D GPE, but from a density estimation.
Therefore, the 2D condensate wave function still remains unsolved.

	In this Letter, we first derive the ``correct" 2D GPE 
from the first principles of two-dimensional scattering theory 
and then show the 2D BEC directly by calculating the condensate wave 
function of trapped atoms.
	The 2D time-independent Schr\"{o}dinger equation 
with the interaction potential $U(\rho)$ is given by
\begin{equation}
(\nabla^2_\rho + k^2 ) \psi_k(\bfrho) = U(\rho) \psi_k(\bfrho),
\label{1}
\end{equation}
where $k^2=2\mu_m E/\hbar^2$.
Here, $E$ is the energy and $\mu_m (=M/2)$ is the reduced mass of two 
identical particles.

The general solution of Eq. (\ref{1}) can be obtained in terms of the 2D 
Green's function such that 
\begin{equation}
(\nabla^2_\rho + k^2 ) G_k(\bfrho,\bfrho') = \delta^2({\bf\rho} - \bfrho') \; .
\label{3}
\end{equation}
Then, the  2D wave function can be expressed as
\begin{eqnarray}
\psi_k(\bfrho) &=& e^{i\bfk\cdot\bfrho} + \int d^2\rho'\,
 G_k(\bfrho,\bfrho') U(\rho')\psi(\rho')
\nonumber \\
&=&  e^{i\bfk\cdot\bfrho} - \frac{i}{4}\int d^2\rho' \,
 H_0(k|\bfrho - \bfrho'|) U(\rho')\psi(\rho')
\nonumber \\ 
&\simeq& e^{i\bfk\cdot\bfrho} 
- \frac{i}{2} \frac{e^{i(k \rho -\frac{\pi}{4})}}{\sqrt{2\pi k \rho}}
\int d^2\rho' \, U(\rho') e^{i(\bfk-\bfk')\cdot\bfrho'} \; .
\label{7}
\end{eqnarray}
Here  $H_0$ is the first-kind Hankel function of order zero 
 defined by $H_0(x)=J_0(x) + i N_0(x)$, where
  $J_0$ ($N_0$) is the Bessel (Neumann) function of order zero.

For large $\rho$, it shows the asymptotic behavior of
$H_0(x) \simeq \sqrt{2/\pi x}\, e^{i(x-\pi/4)}$. 
  Note that  the Born approximation was used and $\bfk' = k \hat{\bfrho}$.
\begin{equation}
\psi_k(\bfrho) \simeq e^{i\bfk\cdot\bfrho} +
F_k \frac {e^{i(k \rho -\frac{\pi}{4})}}{\sqrt{\rho}} \; ,
\label{9}
\end{equation}
where $F_k$ is the  first-order Born scattering-amplitude 
in 2D with a dimension of [length]$^{1/2}$.  Note that 
 this asymptotic formula is a solution of Eq. (\ref{1}).
	Since the scattering cross-section in 2D should have
 the dimension of [length], we can find 
\begin{equation}
\frac{d \sigma_{2D}}{d\theta} = 
\frac{|\bfj_{sc}| \rho}{|\bfj_{in}|} = |F_k|^2,
\label{11}
\end{equation}
where $\bfj_{in}$  ($\bfj_{sc}$) is the incident (scattered) flux density.
Note that the total scattering cross-section for small $k$
becomes $2\pi |F_k|^2$ in 2D  instead of $4\pi |f_{3D}|^2$ in 3D
($f_{3D}$ is the 3D scattering amplitude). 
	Now, if one assumes a delta-function
 type interaction such that  $U(\rho)=(M U_0/\hbar^2) \delta^2(\rho)$
in Eq. (\ref{7}), the $F_k$  can  be written as the following complex form
\begin{equation}
F_k = -\frac{i}{2}\frac{1}{\sqrt{2\pi k}}\frac{M U_0}{\hbar^2} \; . 
\label{13}
\end{equation}

The next step is to find the relation between the scattering 
amplitude  and  scattering length in 2D.   Since the  well-known 
relation  $f_{3D} = -a$, where $a$ is the s-wave scattering length in 3D,
 is not directly applicable to 2D, we need to obtain the relation from 
a partial wave analysis of the two-dimensional  collision theory.
	Outside the range of the potential, the scattered wave 
has a phase shift $\delta_m$, and the scattering matrix becomes
$S_m = e^{2i\delta_m}$ (note that $|S_m|=1$ for elastic scattering).
Therefore, for large $\rho$, the solution of Eq. (\ref{1})
can be written as
\begin{eqnarray}
\psi_k(\bfrho) &=& e^{i\bfk\cdot\bfrho}+ \frac{1}{2}\sum_{m=-\infty}^{\infty}
i^m  (S_m-1) H_m(k \rho)  e^{i m \theta}
\nonumber \\
%%% for single column, choose below   %%%%%%%
%&\simeq & e^{i\bfk\cdot\bfrho}+  \frac{1}{\sqrt{2 \pi k \rho}}
%\sum_{m=-\infty}^{\infty} i^m (e^{2i\delta_m}-1)
%e^{i(k\rho - \frac{m\pi}{2} -\frac{\pi}{4})} e^{i m \theta},
%%% for double column, choose below   %%%%%%%
&\simeq & e^{i\bfk\cdot\bfrho}+  \frac{1}{\sqrt{2 \pi k \rho}}
\sum_{m=-\infty}^{\infty} i^m\times
 \nonumber \\
 &\,& (e^{2i\delta_m}-1)
e^{i(k\rho - \frac{m\pi}{2} -\frac{\pi}{4})} e^{i m \theta},
\label{16}
\end{eqnarray}
where $m$ is an integer.

The scattering length is usually defined as the distance 
for which the two-body wave function is cut off at zero energy. 
 Therefore, the phase shift due to scattering by a potential in 2D
can be expressed in terms of the scattering length so that  \cite{verh}
\begin{equation}
 \delta_0  = \frac{\pi}{2} \frac{1}{\ln k a} \; ,
\label{163}
\end{equation}
where $0 < k a \ll 1$.  Note that $\delta_0 = - k a$ in 3D.
	Eq. (\ref{163}) can be easily obtained by the following simple argument.
The scattering length is an equivalent hard-disk radius $a$,
imposed by the boundary condition $\psi_k(a) = 0$
for the asymptotic wave-function.
For $m=0$, the wave function can then be expressed as, at large distance
and for small $k$ 
\begin{eqnarray}
\psi_k(\bfrho)&=&J_0(k \rho) - \frac{J_0(k a)}{N_0(k a)} N_0(k \rho)
\nonumber \\
&\simeq& \frac{2}{\sqrt{2 \pi k \rho}}
\left[\cos \left(k\rho - \frac{\pi}{4}\right)
- \frac{\pi}{2 \ln ka} \sin \left(k\rho - \frac{\pi}{4} \right)\right]
\nonumber \\
& \equiv &  \frac{2}{\sqrt{2\pi k \rho}}
\cos\left(k\rho -\frac{\pi}{4} +\delta_0\right) \; .
\label{161}
\end{eqnarray}
Note that a necessary condition for the validity of the Born approximation
is that the phase shift $\delta_0$ be very small for small $k$, which can be
easily confirmed  from Eq. (\ref{163}).
Note also that unlike the 3D case where $a$ can be negative as well,
 we do not consider the negative scattering length
since the centrifugal potential of the lowest partial wave is negative in 2D 
[see Eq. (\ref{29})] 
so that the extrapolated local wave function cuts the radial axis 
always above the origin (i.e., $a > 0$) \cite{verh}.

Now, from Eqs.  (\ref{7}),  (\ref{9}) and (\ref{16}), we can express
the scattering amplitude as a series expansion, and also find that 
$m=0$ term is a dominant contribution to the 2D system.  
Therefore,  the 2D scattering amplitude becomes a complex form, given by
\begin{eqnarray}
F_k &=& \frac{1}{\sqrt{2\pi k}}
 \sum_{m=-\infty}^{\infty} \left( e^{2i\delta_m}-1 \right) e^{im\theta}
\nonumber \\
 & = & \frac{2i \delta_0}{\sqrt{2\pi k}} (1 + i \delta_0 + ...)
\nonumber \\
 & \approx & \frac{i \pi}{\sqrt{2\pi k}}\frac{1}{\ln k a} \; .
\label{17}
\end{eqnarray}
	Finally, we can obtain the 2D  interaction potential $U_0$
from Eqs. (\ref{13}) and (\ref{17}) as
\begin{equation}
U_0 =  - \frac{2 \pi \hbar^2}{M}\frac{1}{\ln k a}\; .
\label{201}
\end{equation}
Here, assuming the system is in the ground state of 2D 
harmonic potential, we may approximate the small $k$ to be  $1/a_{ho}$,
where $a_{ho} = \sqrt{\hbar/M\omega}$ ($\om$ is the trap frequency).

Eq. (\ref{201}) is a key result of  this Letter, and 
there will be no dimensional inconsistency in deriving the 2D GPE. 
Now, the 2D condensate wave function can be obtained from the 2D GPE 
such that  \cite{dalf}
\begin{equation}
\left[ -\frac{\hbar^2}{2m}\nabla^2_\rho + V_{ext}(\bfrho)
+ N U_0 \phi^2(\bfrho) \right] \phi(\bfrho)= \mu \phi(\bfrho) \; ,
\label{21}
\end{equation}
where   $\int d^2\rho \, \phi^2(\rho) = 1$
and $N$ is the number of condensate particles.
We assume a 2D isotropic,  harmonic trap potential 
$V_{ext}(\rho) = \frac{1}{2} M \om^2 \rho^2$.
Then we can simplify Eq. (\ref{21}) for numerical calculation
by introducing the dimensionless variables 
($\bfrho \rightarrow a_{ho} \bfrho,$ $ \mu \rightarrow \hbar\om\mu$
 and $\phi \rightarrow a_{ho}^{-1}\phi$) and  the dimensionless interaction 
of $N$ atoms,
$\tilde{U}(N) =  4 \pi  N/|\ln(a/a_{ho}) |$, as
\begin{equation}
\left[  -\nabla^2_\rho +\rho^2 -2\mu +
  \tilde{U} \phi^2(\rho)\right] \phi(\rho)=0 \; ,
\label{27}
\end{equation}
or
\begin{equation}
- \frac {{\rm d}^2 u}{{\rm d}\rho^2} 
+ \left[\rho^2 - \frac{1}{4\rho^2} - 2u 
+ \tilde{U} \frac {u}{\rho}\right]u(\rho) =0 \; ,
\label{29}
\end{equation}
where $\phi(\rho) = u(\rho)/\sqrt{\rho}$. 
Here the 2D chemical potential $\mu$ can be obtained from
 the normalization condition. 
Note that the scattering length $a$ is the only atomic parameter
that contributes to the condensate state.

The procedure to solve  Eq. (\ref{27}) or  Eq. (\ref{29}) is very
similar to the case of 3D \cite{edwa}.
For the non-interacting case, the solution is still Gaussian with
$\phi(\rho) = \pi^{-1/2} e^{-\rho^2/2}$.
In the strongly repulsive limit (Thomas-Fermi limit), 
%on the other hand,
we obtain the parabolic solution
$\phi^2(\rho) = (2\mu - \rho^2)/\tilde{U}$.
With a typical value of $ka \sim a/a_{ho} =  10^{-3}$ in Eq. (\ref{201}),
we have plotted the 2D condensate wave function versus
$\rho$ and $N$ in FIG. 1.  Note that although the 2D
condensate wave functions show the overall behaviors similar to
those of the 3D case, they approach the parabolic limit more rapidly 
with increasing the number of atoms. 
In other words, the effect of atomic interaction potential 
becomes more prominent in 2D.
 Note also that the 2D condensation is not very sensitive to
 the scattering length due to its logarithmic dependence. 
This implies that each bosonic alkali atom with positive scattering 
length may exhibit similar condensate characteristics in 2D.
 Refer to TABLE 1  for more detailed comparisons 
between our 2D results and well-known 3D ones.

The ground-state energy of the 2D system can be obtained from the
energy functional $E$ as \cite{dalf}
\begin{equation}
E[\phi] = \int d^2\rho \left[ \frac{\hbar^2}{2M}|\nabla \phi|^2
+ \frac{1}{2} M\omega^2 \rho^2 |\phi|^2 + \frac{U_0}{2} |\phi|^4\right].
\label{51}
\end{equation}
	With the Gaussian trial function
 $\phi(x) = \sqrt{N/\pi x^2} \,e^{-\rho^2/2x^2}$,  
Eq. (\ref{51}) satisfies the following relation 
\begin{equation}
\frac{E}{N} \geq \sqrt{1 + \frac{N}{|\ln  k a|}} \, \hbar \omega \; .
\label{53}
\end{equation}
The ground-state energy per particle of Eq. (\ref{53}) is plotted in FIG. 2 with
 the same parameter $k a \sim a/a_{ho} =  10^{-3}$ and is compared 
with the well-known 3D result of  $^{87}$Rb.
It is interesting to note that as the number of atom is increased, the 2D system 
becomes much unstable in contrast to the 3D case.

Now, let us consider the vortex states in 2D in connection with 
superfluidity of  the hydrodynamic theory.
The 2D system may rotate around the 2D trap center,
resulting  in a quantized circulation of atoms. 
An angular momentum quantum number $\kappa$ can be then 
assigned to the quantum winding of the 2D vortex state, 
which can be written as 
$\phi(\bfrho) = \psi(\bfrho) e^{iS(\bfrho)}$,
where $\psi(\bfrho) = \sqrt{n(\bfrho)}$ is the modulus. 
When the phase $S$ is chosen as 
$\kappa \theta$ ($\kappa$ is an integer), 
one finds vortex states with a tangential velocity 
$v = \kappa \hbar/M\rho$. 
As a result of the  quantum circulation, the angular momentum
 of the system  with respect to the $\rho = 0$ axis 
becomes  $L = N \kappa \hbar$. 
Including the vortex term in  Eq. (\ref{21}), we now obtain the 2D GPE
with vortex states as
\begin{equation}
\left[  -\nabla^2_\rho + \frac{\kappa^2}{\rho^2} + \rho^2  - 2 \mu +
  \tilde{U} \psi^2(\rho)\right] \psi(\rho)= 0 \; .
\label{37}
\end{equation}
The wave functions for $\kappa = 1$ vortex state are plotted in FIG. 3.
We observe, in particular, that the overall shape and the $N$-dependence of
 the vortex-state wave functions are qualitatively similar to those of 3D \cite{dalf}.
	 Note that 2D BEC transition looks analogous to the 
Kosterlitz-Thouless (KT) vortex-state transition. However, 
the phase transition of 2D BEC does not necessarily involve any interaction 
between atoms and this is the fundamental difference between
the two transitions in 2D.  

Although there are fundamental differences between our purely 2D results 
and 3D ones as  summarized  in TABLE 1, it will be interesting to consider 
the quasi-2D or (2$+$1)D scheme as a limiting case of 3D,
in the sense that the atoms obey 2D statistics while their interactions are 3D.
To compare the situation of 2D trap with an extremely squeezed 3D trap,
we take the 3D external potential as
 $V_{ext}(r_\bot,z) = (1/2) M \omega^2 ( r_\bot^2 + \lambda^2 z^2)$.
Fixing the number of atoms and  increasing the $\lambda$, 
 (note that as $\lambda \rightarrow \infty$, $V_{ext}$ approaches 2D trap).
Solving the 3D GPE with fixed number of atoms,  we have presented
 how the condensate wave functions behave with $\lambda$ in FIG. 4.
We can observe the 3D  wave functions merge to the 2D limit but slowly
as $\lambda$ is increased.

In summary, we have developed the 2D collision theory to obtain
the 2D scattering length and the interaction strength.
We  then derived the 2D nonlinear Schr\"{o}dinger equation (GPE)
for trapped neutral Bose atoms with and without vortex state. 
The 2D condensate and vortex state wave functions are 
calculated numerically from the 2D GPE.
Despite the apparent differences, we observe the
qualitative behaviors similar to the 3D BEC,
but do not expect 2D BEC for negative scattering length.
The ground state energy per particle is calculated 
and compared with that of the well-known 3D traps.
We also have considered the practical quasi-2D case as an extreme
of very thin 3D trap.

%\acknowledgements
Authors thank Chris Greene, Joseph Macek, and H. Nha for useful discussions.
This work was supported by the Creative Research Initiative of the
Korean Ministry of Science and Technology and by 97-N6-02-01-A8 of STEPI,
 Korean Ministry of Science and Technology.
\vfill
\footnotesize{$ ^{\dag}$ E-mail: whjhe@phya.snu.ac.kr}
%%%%%%%%%%%%%%%%%

%%%%%%%%%%%%%%%
\end{multicols}
\newpage
%%%%%%%%%%%%   TABLE   %%%%%%%%%%%%%%%%^
\begin{table}
\caption{ The comparison of the 3D  and  2D BEC.}
\end{table}
%%%%%%%%%%%
\begin{center}

\begin{figure}
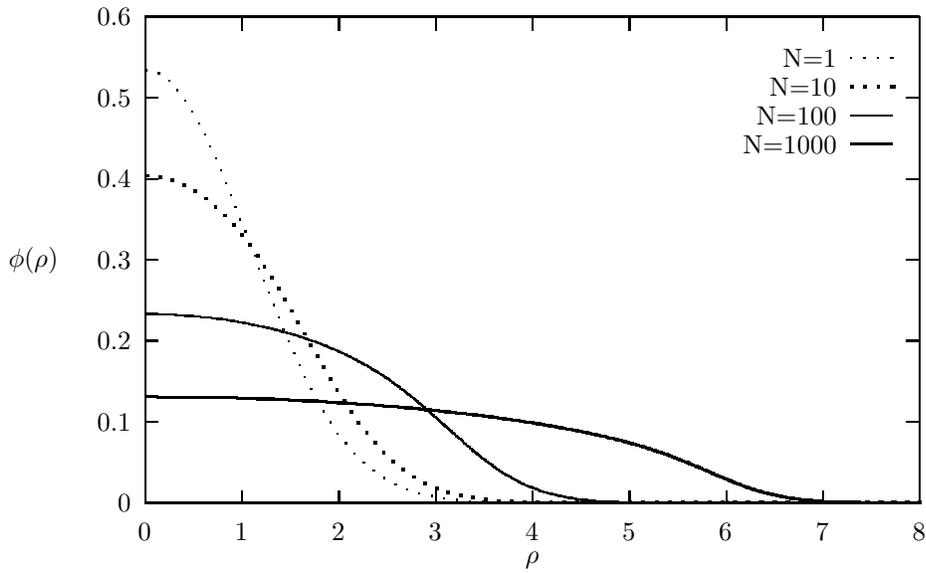

\caption{  The 2D condensate wave functions
 for a typical value of $ka  =10^{-3}$. }
\end{figure}

%%%%   FIGURE 2.  
% GNUPLOT: LaTeX picture
\setlength{\unitlength}{0.240900pt}
\ifx\plotpoint\undefined\newsavebox{\plotpoint}\fi
\sbox{\plotpoint}{\rule[-0.200pt]{0.400pt}{0.400pt}}%
%
\begin{figure}
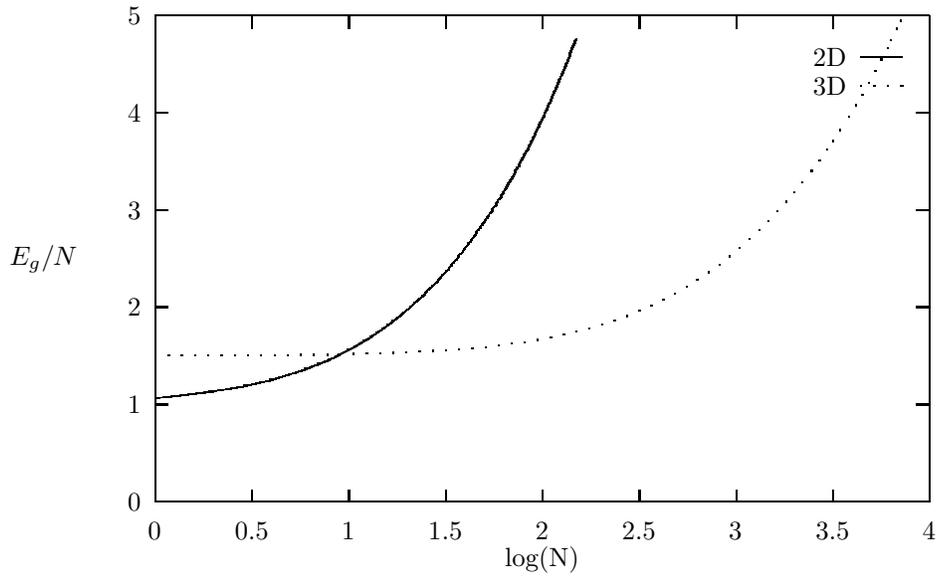

\caption{ The ground-state energy per particle of Eq. (16).
 The unit of y-axis is $\hbar \omega$.}
\end{figure}

%%%%   FIGURE 3. 
% GNUPLOT: LaTeX picture
\setlength{\unitlength}{0.240900pt}
\ifx\plotpoint\undefined\newsavebox{\plotpoint}\fi
\sbox{\plotpoint}{\rule[-0.200pt]{0.400pt}{0.400pt}}%
%
\begin{figure}
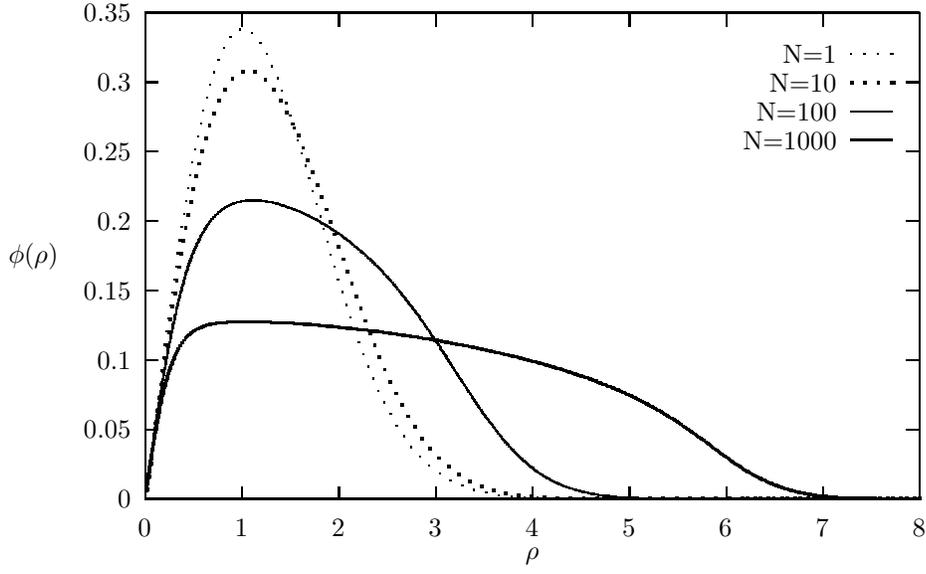

\caption{  The 2D wave functions of vortex state for a quantum 
circulation $\kappa =1$ and a typical value of $k a  =10^{-3}$.}
\end{figure}

%%%%   FIGURE 4.  
% GNUPLOT: LaTeX picture
\setlength{\unitlength}{0.240900pt}
\ifx\plotpoint\undefined\newsavebox{\plotpoint}\fi
\sbox{\plotpoint}{\rule[-0.200pt]{0.400pt}{0.400pt}}%
%
\begin{figure}
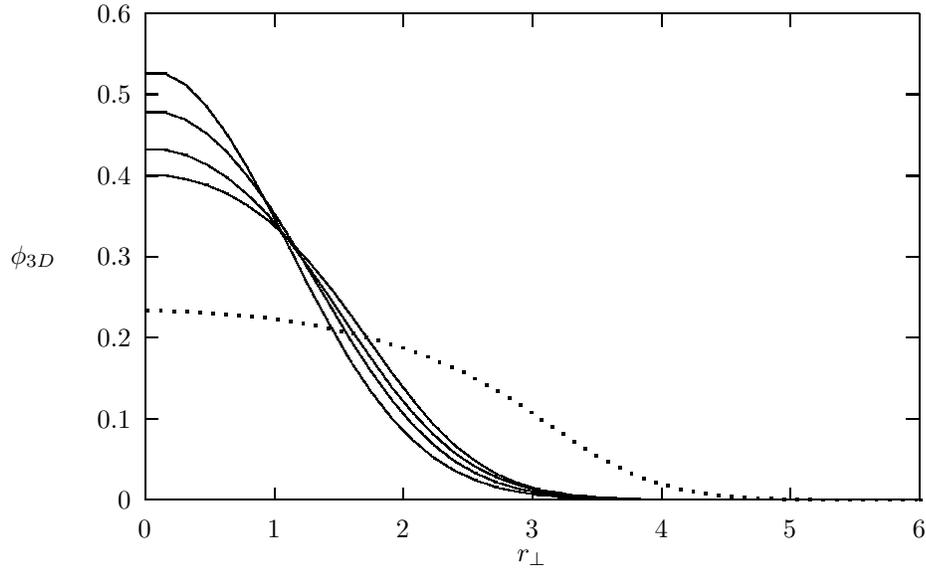

\caption{  The behavior of asymmetric 3D condensate wave functions
for  fixed  $N=100$.  From the top left $\lambda = 10, 100, 300, 1000$.
The dotted line is the 2D condensate wave function of the same $N$. 
Note that $ \phi_{3D} = \left[\int dz |\phi_{3D}(r_{\bot},z)|^2\right]^{1/2}$.}
\end{figure}

\end{document}